\documentclass[aps,preprint]{revtex4}%
\usepackage{amsmath}
\usepackage{graphicx}
\usepackage{amsfonts}
\usepackage{amssymb}%
\providecommand{\U}[1]{\protect\rule{.1in}{.1in}}
\providecommand{\U}[1]{\protect\rule{.1in}{.1in}}

\begin{document}
\title{Quadrupole Plasmon Excitations in Confined One-dimensional Systems}

\begin{abstract}
The existence and nature of a new mode of electronic collective excitations
(quadrupole plasmons) in confined one-dimensional electronic systems have been
predicted by an eigen-equation method. The eigen-equation based on the
time-dependent density-functional theory is presented for calculating the
collective excitations in confined systems. With this method, all modes of
collective excitations in the 1D systems may be found out. These modes include
dipole plasmons and quadrupole plasmons. The dipole plasmon mode corresponds
to the antisymmetric oscillation of induced charge, and can be shown as a
resonance of the dipole response. In the quadrupole plasmon modes, the induced
charge distribution is symmetric, and the dipole response vanishes. The motion
of the electrons in the quadrupole modes is similar to the vibration of atoms
in the breathing mode of phonons. This type of plasmons can be shown as a
resonance of the quadrupole response, and has to be excited by al non-uniform field.

\end{abstract}
\author{Reng-lai Wu, Yabin Yu}
\email{apybyu@hnu.edu.cn.}
\author{Hong-jie Xue}
\affiliation{School of Physics and Microelectronic Science, Hunan University, Changsha
410082, China}

\date{\today}
\maketitle

Plasmon properties in nano-structure systems have\ attracted more and more
physics researcher's attentions, due to their fundamental significance
\cite{1,2,3,4,5,6,7-1,7,8,9,10} and potential applications
\cite{11,12,13,14,15,16,17,18,19,20}. Different from the bulk and
surface-plasmon waves, nanostructures sustain localized plasmon resonances
within their confining boundaries, leading to dynamic charge accumulation and
field strongly enhancement near their surfaces. Such the plasmon oscillations
and the decay at surfaces are responsible for the novel applications in
optical imaging\cite{14}, single-molecule sensing and spectroscopy\cite{15,16}%
, photocatalytic reactions\cite{17}, nano-photonics and -electronics\cite{18}
and cancer therapy\cite{20}.

Collective excitations in few-atom systems shown initially by Kummel et al.
\cite{c1} have been the subject of many theoretical investigations following
recent scanning tunneling microscope observations showing development of 1D
band structure when the number of atoms in Au chains on NiAl(110) exceeds 10
\cite{c2}. Many subsequent theoretical calculation\cite{7-1,7,8,9,10,21,22,23}
confirmed the presence of the collective plasmon mode in the confined
one-dimensional electronic systems of a few atoms. Theoretical studies of
plasmon excitations are mostly done via calculating the dipole response
\cite{7,8,9,10} and other characteristic responses \cite{21,22,23} under
applying an external field, and the excitations are indicated by the
corresponding response resonances. One may wonder whether the modes predicted
in this way are dependent on the applied external fields. The answer is
clearly yes. It has been demonstrated in Ref\cite{8} that a longitudinal field
induces longitudinal-mode plasmon resonance, and a transverse field induces a
transverse resonance. Therefore, finding a proper theoretical approach to
calculate plasmon spectra is essential for further studies of plasmons in
confined electronic systems. In this letter, we attempt to present an eigen
equation of electronic collective-oscillation in confined systems, and use the
eigen equation to find all plasmon excitations of the systems by solving the
equation, and then compare the eigen plasmon excitations with the ones
obtained by the way mentioned above to check whether any new collective
excitation exists in the systems. Consequently, we got a new mode of
electronic collective excitations, quadrupole plasmons, in the confined 1D
electronic systems. In the 1D systems, the dipole plasmon mode corresponds to
the antisymmetric charge oscillation and can be displayed as a resonance of
the dipole response\cite{7,8,9,10}. Distinct from the dipole plasmons, the
quadrupole plasmon corresponds to the symmetric charge oscillation and the
dipole response vanishing, but can be shown as a resonance of the quadrupole
response. The motion of the electrons in the quadrupole modes is similar to
the vibration of atoms in the breathing mode of phonons. Our calculations are
done based on the two models: the one-dimensional electron gas and
one-dimensional tight-binding models. It has been shown in Ref\cite{7,8} and
present work that the longitudinal dipole-plasmon resonances in linear atomic
chains predicted by using a confined 1D electron gas model are qualitatively
in agreement with the calculations made for atomic chains by \textit{ab
initio} time-dependent density functional theory. We believe that the new mode
of collective excitations will exist in the atomic chain systems in
Ref.\cite{7,8,9,10,21,22,23}. Thus we expect that this result will prompt
theoretical and experimental investigations for finding new modes of plasmon
excitations in confined low-dimensional systems, and further affect the future
of nanoplasmonic device engineering and nanoscale photochemistry.

We will start with deducing the plasmon eigen equation, based on the
time-dependent density-functional theory (TDDFT). According to TDDFT, the
induced charge density\cite{24}%
\begin{equation}
\rho(\mathbf{r},\omega)=\int d\mathbf{r}^{\prime}\Pi(\mathbf{r,r}^{\prime
},\omega)V(\mathbf{r}^{\prime},\omega), \label{1}%
\end{equation}
where the Kohn-Sham response function, i.e., the density-density response
function of non-interacting electrons with unperturbed density $\rho_{0}%
,$defined by%
\begin{equation}
\Pi(\mathbf{r,r}^{\prime},\omega)=\frac{\delta\rho\lbrack V](\mathbf{r}%
,\omega)}{\delta V(\mathbf{r}^{\prime},\omega)}|_{V[\rho_{0}]}. \label{1-1}%
\end{equation}
Here we have transformed the time-domain into the frequency-domain. In
Eq.(\ref{1}), the perturbation potential is
\begin{equation}
V(\mathbf{r},\omega)=V^{ex}(\mathbf{r},\omega)+V^{in}(\mathbf{r},\omega),
\label{3}%
\end{equation}
where $V^{ex}(\mathbf{r},\omega)$ is external potential, and
\begin{equation}
V^{in}(\mathbf{r},\omega)=\frac{1}{4\pi\varepsilon_{0}}\int d\mathbf{r}%
^{\prime}\frac{\rho(\mathbf{r}^{\prime},\omega)}{\left\vert \mathbf{r-r}%
^{\prime}\right\vert }+\int d\mathbf{r}^{\prime}K_{xc}(\mathbf{r,r}^{\prime
},\omega)\rho(\mathbf{r}^{\prime},\omega) \label{4}%
\end{equation}
is the induced potential. The time-dependent xc kernel is defined
by$K_{xc}(\mathbf{r,r}^{\prime},\omega)=\delta V^{xc}[\rho](\mathbf{r}%
,\omega)/\delta\rho(\mathbf{r}^{\prime},\omega).$ In fact, the density-density
response function defined in Eq.(\ref{1-1}) is the random-phase approximation
(RPA) Lindhard function%
\begin{equation}
\Pi(\mathbf{r,r}^{\prime},\omega)=2e^{2}\sum_{mn}\frac{f(E_{m})-f(E_{n}%
)}{E_{m}-E_{n}-\omega-i\gamma}\psi_{m}^{\ast}(\mathbf{r})\psi_{n}%
(\mathbf{r})\psi_{n}^{\ast}(\mathbf{r}^{\prime})\psi_{m}(\mathbf{r}^{\prime}),
\label{2}%
\end{equation}
where $f(E_{n})$ is the Fermi-function, $\psi_{n}(\mathbf{r})$ is the energy
eigen-function of \ electrons in the unperturbed system, and $E_{n}$ is the
eigen-energy. The unperturbed eigenstates can be obtained using local density
functional theory. Substituting Eq.(\ref{2}) in Eq.(\ref{1}), we have
\begin{equation}
\rho(\mathbf{r},\omega)=2e^{2}\sum_{mn}\frac{f(E_{m})-f(E_{n})}{E_{m}%
-E_{n}-\omega-i\gamma}\psi_{m}^{\ast}(\mathbf{r})\psi_{n}(\mathbf{r})\left[
V_{n,m}^{ex}(\omega)+V_{nm}^{in}(\omega)\right]  , \label{8}%
\end{equation}
where $V_{nm}^{X}(\omega)=\int d\mathbf{r}V^{X}(\mathbf{r},\omega)\psi
_{n}^{\ast}(\mathbf{r})\psi_{m}(\mathbf{r}).$Combining Eq.(\ref{8}) with
Eqs.(\ref{4}), we can obtain the self-consistent equation for $V^{in}%
(\mathbf{r},\omega)$
\begin{equation}
V^{in}(\mathbf{r},\omega)=2e^{2}\sum_{mn}\frac{f(E_{m})-f(E_{n})}{E_{m}%
-E_{n}-\omega-i\gamma}\int d\mathbf{r}^{\prime}K(\mathbf{r,r}^{\prime}%
,\omega)\psi_{m}^{\ast}(\mathbf{r}^{\prime})\psi_{n}(\mathbf{r}^{\prime
})\left[  V_{n,m}^{ex}(\omega)+V_{nm}^{in}(\omega)\right]  , \label{6}%
\end{equation}
where $K(\mathbf{r,r}^{\prime},\omega)=1/4\pi\varepsilon_{0}\left\vert
\mathbf{r-r}^{\prime}\right\vert +K_{xc}(\mathbf{r,r}^{\prime},\omega).$
Multiplying Eq. (\ref{6}) by $\psi_{m^{\prime}}^{\ast}(\mathbf{r}%
)\psi_{n^{\prime}}(\mathbf{r})$ and integrating over the space yields
\begin{equation}
\sum_{mn}\left[  \delta_{m^{\prime}n^{\prime},nm}-M_{m^{\prime}n^{\prime}%
,mn}(\omega)\right]  V_{nm}^{in}(\omega)=\sum_{mn}M_{m^{\prime}n^{\prime}%
,mn}(\omega)V_{n,m}^{ex}(\omega), \label{7}%
\end{equation}
with%
\[
M_{m^{\prime}n^{\prime},mn}(\omega)=2e^{2}\frac{f(E_{m})-f(E_{n})}{E_{m}%
-E_{n}-\omega-i\gamma}\int d\mathbf{r}\int d\mathbf{r}^{\prime}\psi
_{m^{\prime}}^{\ast}(\mathbf{r})\psi_{n^{\prime}}(\mathbf{r})K(\mathbf{r,r}%
^{\prime},\omega)\psi_{m}^{\ast}(\mathbf{r}^{\prime})\psi_{n}(\mathbf{r}%
^{\prime}).
\]
Now one can calculate the collective charge-oscillation (Eq.(\ref{8})) by
solving Eq.(\ref{7}). Setting $V_{n,m}^{ex}(\omega)=0,$ Eq.(\ref{7}) becomes%
\begin{equation}
\sum_{mn}\left[  \delta_{m^{\prime}n^{\prime},nm}-M_{m^{\prime}n^{\prime}%
,mn}(\omega)\right]  V_{nm}^{in}(\omega)=0. \label{10}%
\end{equation}
This is the plasmon eigen-equation we want, and with the equation all the
plasmon excitations of a system may be found and are not dependent on the
applied external fields. It is worth to point that using the symmetry of
$V_{nm}^{in}(\omega)$ and $M_{m^{\prime}n^{\prime},mn}(\omega),$the number of
equations in Eqs.(\ref{7}) and (\ref{10}) may be reduced. Usually, the eigen
states of a confined system may be expressed by real wave-functions, and in
this case Eqs.(\ref{7}) and (\ref{10}) would be consumedly reduced. According
to eigen-equation (\ref{10}), the plasmon excitation energy $\hslash\omega$
can be determined by $A(\omega)=\det\left[  \delta_{m^{\prime}n^{\prime}%
,nm}-M_{m^{\prime}n^{\prime},mn}(\omega)\right]  =0$. However, there would be
not real solutions due to the finite small imaginary part $i\eta.$ In the
practical calculation a small imaginary part $i\eta$\ is necessary, and the
eigen plasmon excitation energy $\hslash\omega$ is obtained by
$\operatorname{Re}[A(\omega)]=0,$with $\operatorname{Im}\left[  A(\omega
)\right]  \sim0$. This implies that spectrum function $\operatorname{Im}%
\left[  1/A(\omega)\right]  $ will show a peak at the plasmon energy
$\hslash\omega.$ Here we want to point that the eigen resolution should be
exactly real when $i\eta=i0^{+}$, and $A(\omega)$ will give an infinite peak
at plasmon frequency. This eigen-equation method not only let us find out all
the plasmon modes of a system, but also greatly reduces the amount of
computation in comparison with the original TDDFT.

Firt we study the plasmon excitation in a quasi-one-dimensional electron gas
(Q1DEG) confined within a quantum well with length of $(N+1)a$ and width of
$2a$, where $a$ is virtual lattice constant and taken as $35$nm in our
calculation. For this model, the unperturbed wave-function is $\sqrt{\frac
{2}{(N+1)a^{2}}}\sin(\frac{n\pi x}{(N+1)a})\sin(\frac{\pi y}{2a})$. With this
model an atomic chain of $N$ atoms may be mimicked. The similar model was
employed by Gao and Yuan\cite{7-1} to study the plasmon excitation of a atomic
chains. Their calculations indicate that in comparison with the pure RPA the
exchange term $K_{xc}(\mathbf{r,r}^{\prime},\omega)$ gives rise to only very
slight shift in plasmon frequency. In present work, we are only interested in
the qualitative investigation of the plasmon excitation, in particularly
finding out all of the collective excitation in the confined systems.
Therefore, we will ignore exchange term $K_{xc}(\mathbf{r,r}^{\prime},\omega)$
to simplify our calculation.

Our calculation shows that some plasmon eigen-modes corresponding to the peaks
of dipole absorption spectra induced by a local uniform field such as
$V^{ext}(x,t)=-xE_{0}e^{-i\omega t}$\cite{7-1}, but for others of the
eigen-modes there is not appearance of the absorption-peak. In Fig. 1(a) we
show an eigen-mode by a peak of the spectrum function $\operatorname{Im}%
\left[  1/A(\omega)\right]  $ at frequency $\omega\approx0.3058$, and in Fig.
1(b) one can find that this frequency is the zero-point of $\operatorname{Re}%
\left[  A(\omega)\right]  $, where the number of atoms $N=12,$and the number
of electrons $N_{e}=12$. The energy (frequency) is normalized by $\pi^{2}%
\hbar^{2}/2m_{e}a^{2}$, and $m_{e}$ is the mass of electrons. In addition,
taking $V^{ext}(x,t)=-xE_{0}e^{-i\omega t}$ as in Ref.\cite{7-1}, we calculate
the dipole strength (absorption spectrum) $P(\omega)=\omega\int
x\operatorname{Im}[\rho(\mathbf{r},\omega)]dxdy$ by using Eq.(\ref{8}) and
(\ref{7}) In Fig. 1(b), one can find that in the dipole strength functions a
peak appears at the same frequency as in Fig. 1(a). In Fig. 2(a) and (b) we
show another eigen plasmon excitation around frequency $\omega\approx0.5392$.
However, in Fig. 2(c) one can find that there is no peak in the dipole
response function. This indicates that the eigen mode of plasmon shown in Fig.
2 cannot be excited by this applied external field, or this eigen-mode is not
a dipolar plasmon. In addition, we also calculate the induced electric field
energy\cite{21} and the charges $\int\left\vert \rho(\mathbf{r},\omega
)\right\vert dxdy$ as the function of external-field frequency, and find no
any resonance to appear at the eigen frequency of 0.5393. So, we believe that
this excitation mode cannot excited with such the applied field.

In order to understand properties of these eigen plasmon excitations and find
out why some modes of them cannot be excited by the applied field, we now
study the charge distribution in the eigenstates of collective oscillation. By
setting $V_{1,2}^{in}(\omega)=1$ we calculate the charge density
$\rho(\mathbf{r},\omega)$ of the eigen-states at the eigen-oscillation
frequency $\omega=0.306$ and $0.5393,$ and show the results in Fig. 3(a) and
Fig. 3(b) respectively. One can find that the charge-distribution of the two
plasmon modes are very different. In Fig. 3(a), as shown in Ref. \cite{7-1},
both the real and imaginary profiles of the charge-density distribution are
antisymmetric and exhibit likely Friedel oscillations across the systems.
However, in Fig. 3(b) the charge-density distribution is symmetric, so in this
mode of collective excitation the motion of the electrons is similar to the
vibration of atoms in the breathing mode of phonons. It is the symmetric
charge-density distribution to cause this mode unable to be excited by a
uniform external field, because in this case the electric dipole moment
vanishes. The symmetric distribution of charge-density would give rise to a
quadrupole moment, corresponding to the quadrupole plasmon mode. The
quadrupole plasmon resonance modes in nanostructures have widely
reported\cite{25,26,27,28,29}. To the best of our knowledge, however, this
type of plasmon excitation has not been reported in the one-dimensional
cluster systems. Here, we want to point out that both the asymmetry and
symmetry of the charge densities originate from the intrinsic property of the
eigenstates, unlike the argument in Ref. \cite{7-1} that the antisymmetric
charge-density results from the antisymmetric external field. In fact, our
calculations show that the plasmon excitations of antisymmetric charge-density
can be excited by both external fields $V^{ext}(x,t)=-xe^{-i\omega t}$\ and
$V^{ext}(x,t)=-x^{2}e^{-i\omega t}$. However the plasmon excitations of
symmetric charge-density cannot be excited by $V^{ext}(x,t)=-xe^{-i\omega t}$,
but can be excited by $V^{ext}(x,t)=-x^{2}e^{-i\omega t}$.

For the plasmon excitations of symmetric charge-density, the dipolar moment
vanishes, and the interaction energy of the system in uniform applied-field
vanishes, so such the mode of plasmons cannot be excited by the uniform field.
However, the symmetric distribution of charge-density gives the quadrupole
moment $\mathfrak{D}$, and the interaction energy of the quadrupole is
$\mathfrak{D}$:$\nabla\vec{E}$, therefore we predict that the quadrupole mode
of plasmon should be excited by the nonuniform external field. In Fig. 4(a)
and (b) we plot the dipole and quadrupole strength as functions of
external-field frequency, where the external potentials are $V^{ext}%
(x,t)=-E_{0}xe^{-i\omega t}$ and $V^{ext}(x,t)=-F_{0}x^{2}e^{-i\omega t}$ in
Fig. 4(a) and 4(b) respectively. Here, quadrupole strength is defined as
$Q_{e}(\omega)=\omega\int(x-\frac{L}{2})^{2}\operatorname{Im}[\rho
(x,\omega)]dx.$In the dipole response function, one can find the similar
results in Ref. \cite{7-1} and \cite{7}, it is that redshifts in energy and
its intensity increases with the system length. As was pointed in Ref
\cite{7-1} and \cite{7}, the increase in intensity results from the
accumulation of collectivity in the dipole oscillation. The redshift of the
resonance frequency at increased system length can be understood by the
reduction of the energy gaps involved in the dipole excitation. Similar
behaviors are also found in the quadrupole functions, and compared with the
dipole response function, the main plasmon resonance peaks show general
blueshifts for all chain lengths. However, this does not indicate that the
quadrupole plasmon excitations are higher than the dipole plasmons. Due to the
large difference of strength between the different plasmon resonance peaks, in
the plots of Fig. 4 only the contribution of main plasmon peaks can be shown,
and those minor plasmon peaks are invisible. If taking in consideration the
minor plasmon peaks, the lowest is quadrupole plasmon excitation. For
instance, in the system $N=12$ and $N_{e}=12,$ the lowest plasmon excitation
energy is $\omega\approx0.185$, which gives a quadrupole plasmon.

The above investigation is performed based on the model of Q1DEG. In order to
illustrate the generality of the results, we consider the opposite limit case,
namely, we assume that electrons in the atomic chains can be effectively
described using a tight-binding model: the extended Hubbard model,%
\begin{equation}
H=-t\sum_{l<l^{\prime},\sigma}(d_{l\sigma}^{\dagger}d_{l^{\prime}\sigma
}+h.c.)+\sum_{l,\sigma}eV_{l}^{ex}(t)d_{l\sigma}^{\dagger}d_{l\sigma}+\frac
{U}{2}\sum_{l,\sigma}n_{l\sigma}n_{l-\sigma}+\frac{V}{2}\sum_{l\delta}%
n_{l}n_{l+\delta}. \label{12}%
\end{equation}
After a dynamic mean-field approximation, the frequency-dependent charge
response is obtained based on the standard linear response theory,
\begin{equation}
\sum_{l^{\prime}}\left[  \delta_{ll^{\prime}}-\sum_{l^{\prime\prime}}%
\Pi(l,l^{\prime\prime},\omega)v_{l^{\prime\prime}l^{\prime}}\right]  \delta
Q_{l^{\prime}}(\omega)=\sum_{l^{\prime}}e^{2}\Pi(l,l^{\prime},\omega)\left[
V_{l}^{ex}(\omega)\right]  , \label{13}%
\end{equation}
where $\delta Q_{l}(\omega)=e\delta n_{l}(\omega)$ is charge response on site
$l,$ $\Pi(l,l^{\prime},\omega)=2\sum_{mn}\frac{f(E_{m})-f(E_{n})}{E_{m}%
-E_{n}-\omega-i\gamma}\psi_{m}^{\ast}(l)\psi_{n}(l)\psi_{n}^{\ast}(l^{\prime
})\psi_{m}(l^{\prime})$ is the Lindhard function, and $v_{ll^{\prime}}%
=\frac{U}{2}$ for\ $l=l^{\prime}$ and $v_{ll^{\prime}}=V$ for\ $l=l\pm1.$
Here, $E_{n}=-2t\cos(\frac{n\pi}{N+1}),n=1,2...N,$and $\psi_{n}(l)=\sqrt
{\frac{2}{N+1}}\sin(\frac{n\pi}{N+1}l).$ Following from Eq. \ref{13}, both the
eigen plasmon excitations and external field induced resonances can be obtain.
Qualitatively, the eigen excitations show the same behavior as in the Q1DEG
model, and can be divided into two types of plasmons, i. e., dipole and
quadrupole plasmon. We show the dipole and quadrupole response functions for
different chain-length in Fig. 4(a) and 4(b) respectively, where the energy
(frequency) is taken the hopping $t$ as unit, and $U=3,V=1$.

In conclusion, we have presented an eigen-equation method for studying plasmon
excitations, which is based on TDDFT. Using the method, we have studied the
plasmon excitations in the confined Q1DEG systems, and predicted a new type of
plasmon excitations, i. e., the quadrupole plasmons, which has not been
reported in the one-dimensional systems. Different from the dipole plasmon
excitations that correspond to the antisymmetric distribution of induced
charge, the new type of plasmons correspond to the symmetric charge
distribution, indicating that the motion of charges in this mode of plasmon
oscillation is similar to the atomic vibration in the breathing mode of
phonons. Since the dipole moment vanishes in the quadrupole mode of plasmons,
their excitation only can be achieved\ by applying a nonuniform external
field, and the quadrupole plasmons may display as the resonance peaks of
quadrupole response function. Furthermore, we have shown that the quadrupole
plasmon excitations can be predicted in the extended Hubbard models. We expect
that this result will prompt theoretical and experimental investigations for
finding new modes of plasmon excitations in confined systems, and affect
nanoplasmonic device engineering and nanoscale photochemistry.

The work is supported by NSF of China Grants No. 1077404.

\section{FIGURE CAPTIONS}

FIG. 1 An eigen plasmon excitation of dipole-mode in the Q1DEG is shown by a
peak of the spectrum function $\operatorname{Im}\left[  1/A(\omega)\right]  $
(a), and corresponds to a zero-point of the function $\operatorname{Re}\left[
A(\omega)\right]  $ (b). Applying the external potential $V^{ext}%
(x,t)=-E_{0}xe^{-i\omega t}$, a resonance peak of dipole response function (c)
appears at the eigen excitation. Here the number of atoms $N=12,$and the
number of electrons $N_{e}=12$. The energy (frequency) is normalized by
$\pi^{2}\hbar^{2}/2m_{e}a^{2}$, and $m_{e}$ is the mass of electrons.

FIG. 2 An eigen plasmon excitation of the new-mode is shown by a peak of the
spectrum function $\operatorname{Im}\left[  1/A(\omega)\right]  $ (a), and by
a zero-point of the function $\operatorname{Re}\left[  A(\omega)\right]  $
(b). In the plot of the dipole response function (c), there is no resonance
peak for this mode of plasmon. The parameters are the same as in Fig. 1.

FIG. 3 (Color online) The distribution of the induced charge density for the
eigen plasmon excitation of dipole mode (a) and quadrupole mode.

FIG. 4. (Color online) (a) The dipole response of the Q1DEG as a function of
the external-field frequency $\omega,$obtained by applying the external
potential $V^{ext}(x,t)=-E_{0}xe^{-i\omega t}$ for different system length
$L=(N+1)a$. Here $N$ is the number of atoms with interatomic distance
$a=0.35$nm, and $N_{e}$ the number of electrons. (b) The quadrupole response
function obtained by applying the external potential $V^{ext}(x,t)=-F_{0}%
x^{2}e^{-i\omega t}.$Compared with the dipole response function, the main
plasmon resonance peaks show general blueshifts for all system lengths.

FIG. 5. (Color online) (a) The dipole response function of the atomic chains
obtained based on the extended Hubbard model, for different numbers of atoms
$N,$and number of electrons $N_{e}$. (b) The quadrupole response function the
atomic chains obtained based on the extended Hubbard model. Compared with the
dipole response function, the main plasmon resonance peaks show general
blueshifts for all chain lengths. Here the energy (frequency) is taken the
hopping $t$ as unit, and $U=3,V=1$.

\end{document}